\documentclass[12pt,preprint]{aastex}

\shorttitle{Rapid star cluster disruption in starburst rings}
\shortauthors{de Grijs \& Anders}

\begin{document}
\title{Extremely rapid star cluster disruption in high-shear
  circumnuclear starburst rings: the unusual case of NGC 7742}

\author{Richard de Grijs\altaffilmark{1,2} and Peter
  Anders\altaffilmark{1}}

\altaffiltext{1}{Kavli Institute for Astronomy and Astrophysics,
  Peking University, Yi He Yuan Lu 5, Hai Dian District, Beijing
  100871, China; grijs@pku.edu.cn, anders@pku.edu.cn} 
\altaffiltext{2}{2012 Selby Fellow, Australian Academy of Science}

\begin{abstract}
All known mass distributions of recently formed star cluster
populations resemble a `universal' power-law function. Here we assess
the impact of the extremely disruptive environment in NGC 7742's
circumnuclear starburst ring on the early evolution of the galaxy's
high-mass ($\sim 10^5$--$10^7 M_\odot$) star cluster
population. Surprisingly, and contrary to expectations, at all ages --
including the youngest, $\la 15$ Myr -- the cluster mass functions are
robustly and verifiably represented by lognormal distributions that
resemble those commonly found only for old, evolved globular cluster
systems in the local Universe. This suggests that the high-shear
conditions in the NGC 7742 starburst ring may significantly speed up
dynamical star cluster destruction. This enhanced mass-dependent
disruption rate at very young ages might be caused by a combination of
the starburst ring's high density and the shear caused by the
counterrotating gas disk in the galaxy's inner region.
\end{abstract}


\section{Introduction}
\label{intro.sec}

Circumnuclear rings in spiral galaxies represent environmental
conditions that are conducive to intense star and star cluster
formation. Since the early work by S\'ersic \& Pastoriza (1965, 1967),
many intensely star-forming rings featuring compact `hot spots' have
been identified, while high spatial resolution {\sl Hubble Space
  Telescope (HST)} observations have revealed the presence of numerous
young (a few $\times 10^7$ yr) and intermediate-age (up to a few Gyr)
star clusters in these structures (e.g., Barth et al. 1995; Maoz et
al. 1996, 2001; Ho 1997; Buta et al. 2000; de Grijs et al. 2003;
Mazzuca et al. 2008; van de Ven \& Chang 2009; Hsieh et al. 2012).

NGC 7742, an almost face-on, SA(r)b-type spiral galaxy at a distance
$D = 22.2$ Mpc ($1'' \equiv 107$ pc) that features a $\sim 2$
kpc-diameter circumnuclear ring, is an unusual example of this class
of galaxies, however: it does not have a dominant bar (but see Laine
et al. 2002; Comer\'on et al. 2008). Sil'chenko \& Moiseev (2006)
obtained two-dimensional kinematic data (a combination of
integral-field and long-slit spectroscopy) and deep optical imaging of
the galaxy. They interpreted the presence of a {\it global} gaseous
subsystem that is counterrotating with respect to the {\it global}
disk-like rotation of the stars, combined with the starburst ring, as
the result of a past minor merger with a gas-rich dwarf galaxy (see
also Mazzuca et al. 2006; Tutukov \& Fedorova 2006). The
non-axisymmetric gravitational perturbations resulting from such a
merger are similar to those associated with a bar. Both types of
perturbations can result in a torque change at a particular radius
and, hence, accumulation of gas into a circumnuclear ring.

Here we explore the mass distribution of the young star clusters
associated with the galaxy's starburst ring. Our aim is to assess the
impact of the ring's high-shear conditions on the formation and early
evolution of this star cluster population.

\section{Observations and data reduction}

We obtained high-resolution images of NGC 7742 from the {\sl HST} Data
Archive. The galaxy was observed with the Wide-Field and Planetary
Camera/2 (WFPC2) in the F336W, F555W, F675, and F814W filters with
exposure times of 4200, 960, 960, and 1360 s, respectively, as part of
proposal GO-6276 (PI Westphal). The galaxy's center and its
circumnuclear ring were centered on the WF3 chip ($0.1''$
pixel$^{-1}$); all four exposures were aligned to subpixel accuracy
(neither image rotation nor scaling was required). Although we also
explored the use of a near-infrared F160W image obtained with the
Near-Infrared Camera and Multi-Object Spectrometer (NICMOS), the field
of view sampled and the signal-to-noise ratio (S/N) -- combined with
the smoother distribution of old stellar populations traced by
near-infrared light compared to young star-forming regions -- proved
insufficient.

\subsection{Basic reduction}

Our data reduction followed the detailed description and justification
in de Grijs et al. (2012), which we briefly summarize here. The
standard deviations ($\sigma_{\rm sky}$) of the numbers of counts in
empty sections in all images were established to ascertain a `sky'
background count for each filter. We used multiples of this background
count as thresholds above which the numbers of source detections in
both the F555W and F814W filters were calculated. The number of
detections initially decreases rapidly with increasing detection
threshold. This indicates that our `source' detections are
noise-dominated in the low-threshold regime. Where the rapid decline
flattens off, the detections become dominated by `real' objects,
either cluster candidates or background-intensity variations. In view
of the images' individual S/N characteristics, the most suitable
thresholds, at 4 and $3 \sigma_{\rm sky}$, were 0.067 and 0.061 counts
s$^{-1}$ (917 and 764 sources) for the F555W and F814W filters,
respectively.

We subsequently employed a cross-identification procedure to select
the 398 sources that coincided with intensity peaks within 1.4 pixels
of each other in the F555W$\otimes$F814W filter combination (i.e.,
allowing for 1-pixel mismatches in both spatial directions). Next, we
applied a Gauss-fitting routine to each candidate cluster. Although we
do not expect to detect individual stars at the distance of NGC 7742,
size selection can help us to remove unlikely cluster candidates. We
generated model WF3 point-spread functions (PSFs) using {\sc TinyTim}
(Krist \& Hook 2001) and applied a conservative clipping approach to
weed out source detections characterized by Gaussian widths in the
F555W image, $\sigma_{\rm F555W}$, that were too small to be `real'
point sources (i.e., $\sigma_{\rm F555W} \le 0.8$ pix), such as cosmic
rays or detector artefacts. This left us with a sample containing 352
sources. The resulting size distribution exhibits a clear peak near
the PSF size (with $\sim$80\% of the clusters characterized by
$\sigma_{\rm F555W} \la 2$ pixels $\simeq 22$ pc), with a shallow tail
toward more extended objects. We checked that the five most extended
objects ($4 \le \sigma_{\rm F555W} \le 9$ pixels, or $\sigma_{\rm
  F555W} \lesssim 100$ pc) are associated with clear star-forming
complexes. They are unlikely background galaxies, but they might be
cluster complexes or cluster mergers (e.g., Fellhauer \& Kroupa 2005;
Br\"uns et al. 2011).

\subsection{Photometry}

Our custom-written {\sc idl} aperture-photometry task employs source
radii and sky annuli {\it individualized} for each cluster
candidate. We used a source aperture radius of $3\times \sigma_{\rm
  F555W}$, and 3.5 and $5\times \sigma_{\rm F555W}$ for the inner and
outer sky annuli, respectively. This was based on inspection of the
stellar growth curves, to identify where the object profiles disappear
into the background noise. We verified that our source radii were
chosen conservatively and sufficiently far out so as not to exclude
any genuine source flux. We also checked that the background annuli
were chosen appropriately and not dominated by neighboring
sources. Our observations are characterized by a 50\% completeness
limit at $m_{\rm F555W} = 23.0$ (ST) mag for the galaxy as a whole and
$m_{\rm F555W} = 22.0$ mag in the starburst ring. Following common
practice, we quote our completness limit in the $V$-band
filter. However, our analysis is limited by the F336W image, which has
the lowest S/N. The corresponding limits are $m_{\rm F336W} = 22.5$
and 21.5 mag, respectively.

As a final step, we removed objects that lacked a robustly determined
flux in one or more filters: we need a minimum of four filters to
determine reliable cluster parameters (Anders et al. 2004). Our final
sample contained 317 objects. Three quarters of the sources rejected
in this final step lacked reliable photometry in the F336W filter.

\section{Cluster ages and masses}

\subsection{Models}

For our cluster age and mass determinations, we adopted the set of
{\sc galev} simple stellar population (SSP) models (Kotulla et
al. 2009; and references therein) assuming solar metallicity, based on
the good match of previous NGC 7742 analyses to solar-metallicity
models (Mazzuca et al. 2006; Peletier et al. 2007; Sarzi et
al. 2007). For a given cluster, we used the {\sc AnalySED} tools
(Anders et al. 2004) to compare the shape of the observed
spectral-energy distribution (SED) with the shapes of {\sc galev}
model SEDs (as a function of age and foreground extinction). The
associated cluster mass was derived from the magnitude offset between
the observed and model SEDs. Each model SED (and its associated
physical parameters) was assigned a probability based on the $\chi^2$
value of this comparison. The model data set with the highest
probability was adopted as the most representative set of
parameters. Models with decreasing probabilities were summed up until
a 68.26\% (1$\sigma$) total probability was reached, to estimate the
uncertainties in the best-fitting model. These uncertainties are upper
limits, since their determination specifically also takes into account
effects such as the presence of several local minima for a given
cluster (e.g., owing to the age--metallicity degeneracy) and
discretization in parameter space.

\begin{figure}
\includegraphics[width=\columnwidth]{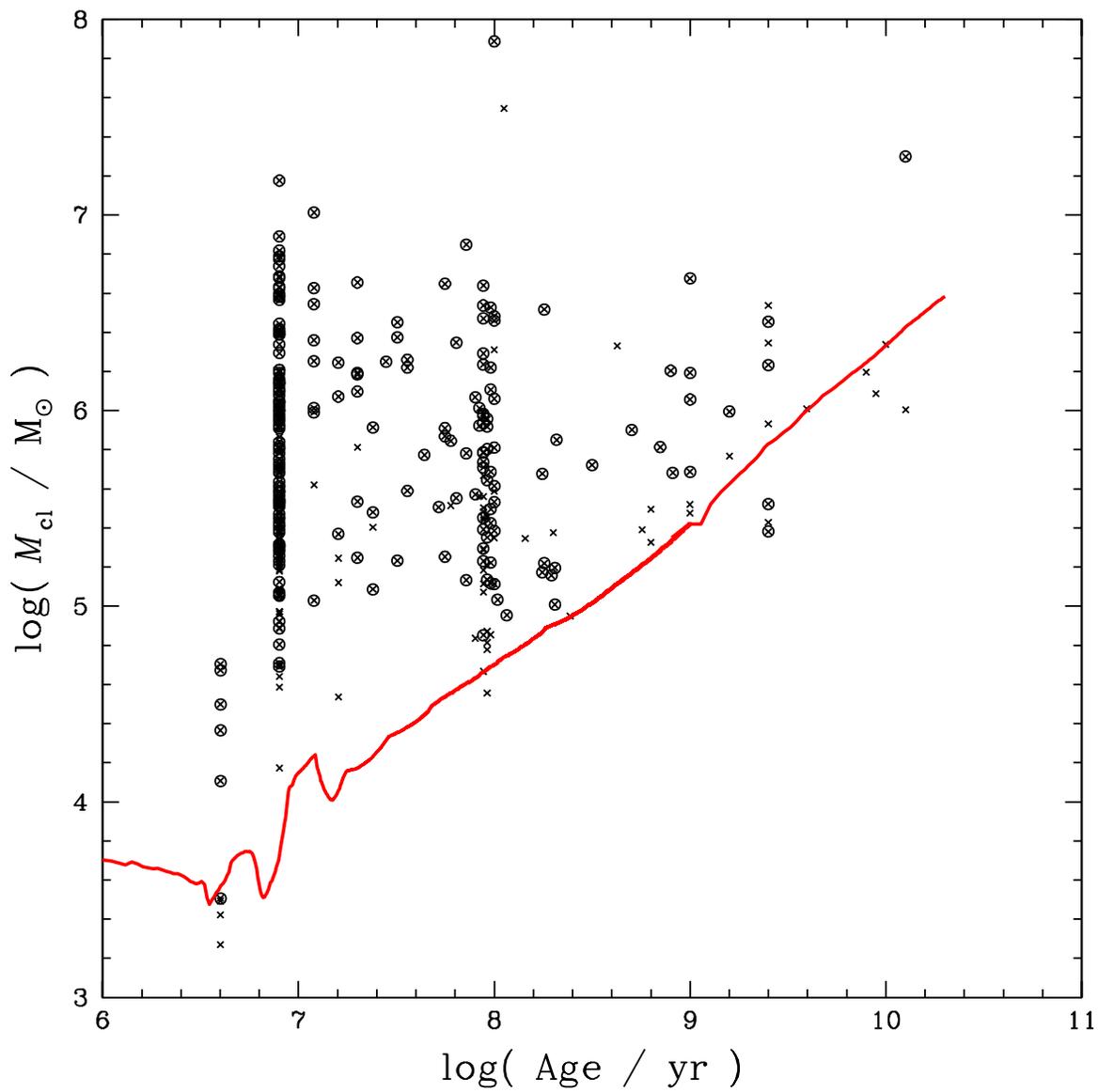}
\caption{\label{agemass.fig}The NGC 7742 star cluster population in
  the diagnostic age--mass plane. Error bars have been omitted for
  reasons of clarity (see text for a discussion of their impact). The
  solid line represents SSP evolution at our 50\% completeness limit,
  adopting $m_{\rm F555W} = 23.0$ mag. Circled crosses: starburst ring
  clusters.}
\end{figure}

Fig. \ref{agemass.fig}, shows the derived ages and masses for our 317
cluster candidates. The solid line represents the expected evolution
of an SSP for $m_{\rm F555W} = 23.0$ mag. The 256 starburst ring
clusters are indicated as circled crosses. We consider clusters to be
part of the ring if they are located at radii, $R$, of $0.52 \le R \le
1.35$ kpc. Adopting the more conservative range of $0.72 \le R \le
1.35$ (Mazzucca et al. 2008) would remove only seven objects from our
sample.

\subsection{Age distribution}

Fig. \ref{agemass.fig} indicates that most clusters have ages of
either around 10 or 100 Myr, but this result is caused by the effects
of discretization. It is well-known that age estimates based on
broad-band imaging observations tend to be narrowly confined to
`chimneys' associated with local minima in parameter space; the
effects of interpolating discrete isochrones for cluster age
determinations and the resulting artifacts in age-–mass space were
discussed in depth by Bastian et al. (2005) and Gieles et
al. (2005). The chimney near 10 Myr corresponds to the appearance of
red supergiants in stellar populations, while 100 Myr is the age at
which asymptotic giant branch stars first show up. 

The majority of clusters in the starburst ring are found in either one
of the two chimneys, roughly in a 2:1 ratio between the younger and
older chimneys. They are generally well mixed along the ring, although
a handful of clusters younger than 10 Myr is found in the northern and
northeastern sections of the ring. This behavior is expected given
that stars and gas tend to circulate along such structures in
typically a few $\times 10^7$ yr (cf. Sarzi et al. 2007). Although the
stellar age map of Peletier et al. (2007) does not provide us with
sufficient spatial information to assess the validity of our results,
they state that the ring is composed of young stellar populations (see
also Sarzi et al. 2007). Mazzuca et al.'s (2008) age distribution of
38 H{\sc ii} regions shows qualitatively similar trends as deduced
here.

\subsection{Masses and mass functions}

Approximately 95\% of our sample clusters have derived masses $\ga
10^5 M_\odot$. Based on our recent analysis of the impact of
stochastic sampling of stellar mass functions (MFs) on integrated star
cluster properties (de Grijs et al. 2012; see also Silva-Villa \&
Larsen 2011) we conclude that the effects of stochasticity are minimal
in the mass range covered by the NGC 7742 clusters. In de Grijs et
al. (2005), we concluded that application of the {\sc AnalySED}
approach based on the {\sc galev} models led to high-accuracy relative
mass determinations within a given cluster system (Anders et
al. 2004). We found that the accuracy with which the cluster mass
distribution can be reproduced using different approaches (including
different SSP models, filter combinations, and input physics) is
$\sigma_M = \Delta \langle \log(M_{\rm cl} / M_\odot ) \rangle \le
0.14$. In fact, we concluded that ``[t]his implies that mass
determinations are mostly insensitive to the approach adopted.'' This
is so, because the mass-to-light ratio of a given SSP depends only
weakly on the population's age.

Since the detection limit of our star clusters is a function of their
age, we need to carefully select the age and mass ranges over which to
compare the cluster population's evolution. In Fig. \ref{mfs.fig} we
compare the MFs of the clusters in the ring as a function of age. The
error bars represent Poissonian uncertainties. We focus specifically
on the two dominant age ranges, i.e., $6.8 \le \log( t / {\rm yr}) \le
7.2$ and $7.8 \le \log( t / {\rm yr}) \le 8.2$. The corresponding mass
ranges that are affected by negligible or only limited sample
incompleteness (up to 50\%) for these two age bins are $\log(M_{\rm
  cl}/M_\odot) \ge 4.0$ and 4.5, respectively, where we have corrected
our analysis for the effects of up to 50\% incomplete sampling. (We
also show the MFs of the galaxy's entire cluster population in the
relevant age ranges as black open circles for comparison.) Given the
extremely young age range of the youngest selection, it is surprising
that the cluster MFs resemble a lognormal distribution in mass. To
verify that this is a robust result, we confirmed that the cluster MF
in the age range between both chimneys exhibits a similar behavior.

\section{Implications and Concluding Remarks}
\label{implications.sec}

\begin{figure}
\includegraphics[width=\columnwidth]{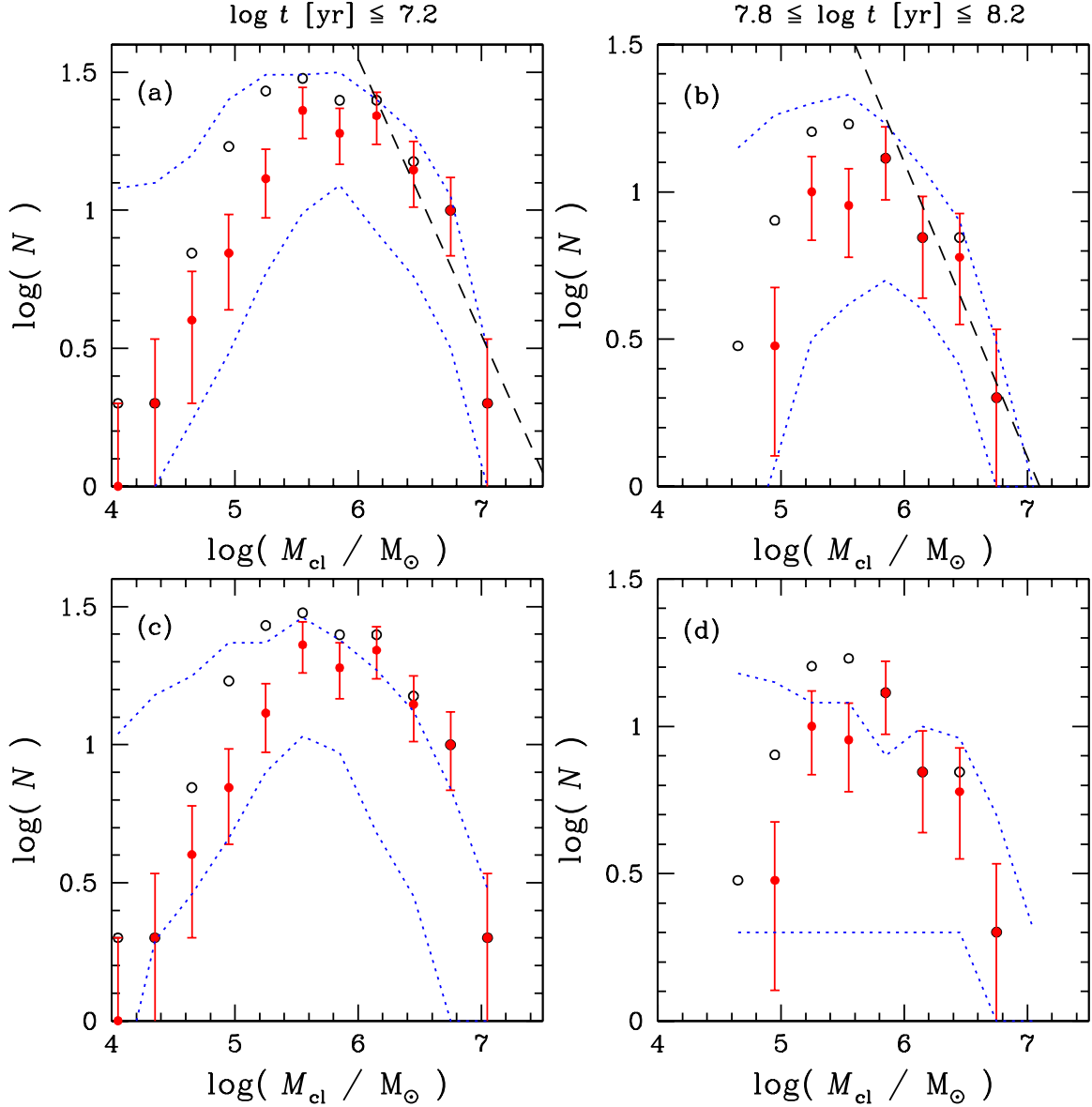}
\caption{\label{mfs.fig}NGC 7742 cluster MFs and their Poissonian
  uncertainties for ages $t$ of $\log( t / {\rm yr}) \le 7.2$
  (left-hand panels) and $7.8 \le \log( t / {\rm yr}) \le 8.2$
  (right-hand panels) and for all clusters in the galaxy above our
  50\% completeness limit (corrected for incompleteness). Red solid
  bullets: Clusters in the starburst ring; black open circles: MFs of
  the full, galaxy-wide cluster populations. The blue dotted lines
  indicate the upper and lower envelopes of the ring-cluster MFs based
  on two statistical redeterminations of the cluster ages and masses
  (top panels: Gaussian error distribution; bottom panels: uniform
  uncertainties; see \S \ref{implications.sec}). The dashed lines
  represent the expected power-law mass distributions with an index of
  $\alpha = 2$.}
\end{figure}

It is now well established that cluster MFs at the time of star
(cluster) formation are well described by power-law distributions of
the form ${\rm d}N/{\rm d}M_{\rm cl} \propto M_{\rm cl}^{-\alpha}$,
where $\alpha$ is usually close to 2 (cf. de Grijs et al. 2003;
Portegies Zwart et al. 2010; Fall \& Chandar 2012). This power-law
form seems ubiquitous and independent of initial stellar or gas
density: it is found in environments ranging from the low-density,
quiescent Magellanic Clouds (de Grijs \& Anders 2006; de Grijs \&
Goodwin 2008) to the high-density, violently interacting Antennae
galaxies (e.g., Whitmore et al. 2010). An initial power-law MF seems,
therefore, a reasonable boundary condition for our discussion of
Fig. \ref{mfs.fig}.

If we assume, not unreasonably given the long timescales involved,
that the highest-mass clusters in NGC 7742 essentially represent their
formation conditions, we can approximate the high-mass end of the MFs
by a power law with a slope of $-2$ (see the dashed lines in
Fig. \ref{mfs.fig}). The approximately lognormal MF shapes imply that
a large fraction of the lower-mass clusters are `missing' compared to
expectations. Based on our current best understanding of the early
evolution of cluster MFs, it is particularly surprising that the
cluster MF in the youngest age bin resembles a lognormal distribution
(in fact, a power-law mass function is generally a poor fit, even for
the highest masses of $M_{\rm cl} \ga 10^6 M_\odot$). Except if we
release the assumption that the initial cluster MF was a power law --
which would contradict most observational studies in this very active
area of current research -- three effects could have caused this
discrepancy: (i) technical issues related to our age and mass
determinations, (ii) evolution of the star cluster population on very
short timescales ($\lesssim 10^7$ yr), and/or (iii) differences in the
cluster formation conditions compared to other environments featuring
large samples of young star clusters. Note that issues related to
sample incompleteness will only affect the lowest cluster masses in
our completeness-corrected samples, i.e., masses {\it significantly
  below} the observed peaks of the MFs.

To ascertain whether the observed `turnover' in the NGC 7742 cluster
MF could have been caused by artefacts related to our fitting
approach, we performed two statistical tests. For each of our sample
clusters we considered the original age determination and its
$1\sigma$ uncertainties. We then randomly drew a new age from within
this uncertainty range and -- using the observed fluxes --
redetermined its mass based on the appropriate, age-dependent
mass-to-light ratio provided by our SSP models. We redetermined the
cluster mass 1000 times using two different probability distributions
in age. The first adopted the highest probablility given by the
original age determination and with a (skewed) Gaussian probability
distribution of which the width and rate of decrease were determined
by the original error bars (blue dotted envelopes in
Fig. \ref{mfs.fig}, top panels). Our second test was based on a
uniform probability distribution within the error range determined for
the original age estimate (Fig. \ref{mfs.fig}, bottom panels). Our
random resampling of the cluster ages and the subsequent
redetermination of their masses firmly rules out an original power-law
MF. This implies that the derived MF shape is robust with respect to
artefacts introduced by our fitting approach. (Note that the
systematic offset in mass for the resampling test based on a uniform
probability distribution reflects the systematically larger
uncertainty ranges toward older ages compared to the error bars toward
younger ages.)

It thus appears that at least part of the observed `downturn' toward
lower masses may be due to either the effects of star cluster
evolution or their formation conditions. From an evolutionary
perspective, the conditions in the galaxy's starburst ring appear to
speed up the destruction or evaporation of a large fraction of the
lower-mass clusters. We speculate that this enhanced cluster
disruption rate at very young ages is caused by a combination of the
high stellar and gas density in the starburst ring\footnote{If we
  adopt Mazzucca et al.'s (2008) ring size and assume that the ring
  thickness follows that of generic, late-type galactic disks, with a
  ratio of scale length to scale height of order eight (e.g., de Grijs
  1998), we obtain an approximate density of 300 massive clusters
  kpc$^{-3}$ or an average separation of $\sim 150$ pc. Note that
  crowding of clusters in the ring is not an issue for our cluster
  detection routines and subsequent photometry.} and the shear caused
by the galaxy-wide counterrotating gas disk. Although the gas and
stellar disks rotate in opposite senses globally, the high gas density
and filling factor (cf. Falc\'on-Barroso et al. 2006) and our
detection of very young clusters in the ring (which presumably formed
{\it in situ} and are hence expected to co-rotate with the gaseous
subsystem) may have led to increased shear, as well as star/gas and
star (cluster)/star (cluster) interactions in the ring. In addition,
at the ring's galactocentric radius, the galaxy's rotation curve has
already reached its constant level (Sil'chenko \& Moiseev 2006), so
that differential rotation across the ring will induce additional
shear.

Note that the peak of the cluster MF occurs near $\log( M_{\rm
  cl}/M_\odot) = 5.5$--5.6, i.e., close to the `universal' peak of the
(old) globular cluster MF in the Milky Way and nearby galaxies, $\log(
M_{\rm cl}/M_\odot) \simeq 5.3$. Stellar evolution of a
solar-metallicity SSP from 10 Myr to 10 Gyr will lead to a reduction
in mass by $\sim 35$--40\% due to stellar mass loss, i.e., through
stellar winds and supernova activity (cf. Leitherer et al. 1999). We
thus anticipate that continued cluster disruption, combined with
evolutionary mass loss, will result in an even closer match by the
time the young ring clusters reach similarly old ages.

The only alternative explanation of the derived lognormal MF is that
the high shear in the starburst ring has created an environment in
which the formation of low(er)-mass clusters ($M_{\rm cl} \lesssim
\mbox{ a few} \times 10^4 M_\odot$) is suppressed, while high-mass
clusters are formed more readily. However, this seems contrary to
observational results, where more massive star clusters are usually
found in environments affected by less shear (e.g., Weidner et
al. 2010). On the other hand, Dowell et al. (2008) concluded that
``[t]he observed similarity of the initial cluster mass functions in
irregular and spiral galaxies implies that the process determining the
masses of clusters does not depend on galactic shear'' (see also Zhang
et al. 2001; Whitmore et al. 2005). We thus conclude that the high
shear in the NGC 7742 circumnuclear starburst ring has most likely led
to extremely rapid cluster disruption, a view corroborated by the
numerical simulations of young clusters in the similarly forbidding
environment of the Galactic Center (e.g., Kim et al. 1999; Portegies
Zwart et al. 2001). Further supporting (although circumstantial)
evidence is found in the MF of the objects in the youngest chimney
that are located outside the starburst ring. Although this sample
includes only 45 objects, it exhibits a clear power-law MF for $M_{\rm
  cl} \ga 10^5$ M$_\odot$ (i.e., a power law extending to
significantly lower masses than for the clusters in the ring, by up to
an order of magnitude), indicative of a dynamically younger cluster
population.

\section*{Acknowledgements} 
We thank Doug Lin for useful discussions. This paper is based on
archival observations with the NASA/ESA {\sl HST}, obtained from the
ST-ECF archive facility. This research has also made use of NASA's
Astrophysics Data System Abstract Service. We acknowledge research
support through grant 11073001 from the National Natural Science
Foundation of China.

\end{document}